\documentstyle[12pt]{article}

\def\be{\begin{equation}}
\def\ee{\end{equation}}
\def\bea{\begin{eqnarray}}
\def\eea{\end{eqnarray}}

\def\br{\begin{thebibliography}{abc}}
\def\er{\end{thebibliography}}
\def\rf{\bibitem}

\def\a{\alpha}
\def\b{\beta}

\def\d{\delta}
\def\e{\epsilon}
\def\f{\phi}

\def\s{\sigma}

\def\D{\Delta}

\def\G{\Gamma}

\def\O{\Omega}

\def\rt{\rightarrow}

\def\bar#1{\overline{#1}}

\def\Hat#1{\rlap{\kern.10em$\widehat{\phantom G}$}#1}
\def\HAt#1{\rlap{\kern.05em$\widehat{\phantom G}$}#1}

\def\czp#1{\rlap{\kern.1em$\widehat{\phantom{G\vrule height.8em}}$}#1{}}
\def\Czp#1{\rlap{\kern.05em$\widehat{\phantom{G\vrule height.8em}}$}#1{}}

\def\I1{\relax{\rm 1\kern-.28em l}}
\newcommand{\sect}[1]{\setcounter{equation}{0}\section{#1}}

\def\sxn#1{\bigskip\medskip \sect{#1} \smallskip
                                                 }

\begin{document}

\thispagestyle{empty}
\setcounter{page}{0}

\begin{flushright}
IFUSP-P- 1246
\end{flushright}

\vspace{1cm}

\begin{center}
{\LARGE Quasi-Topological Field Theories in Two Dimensions \\
\vspace{5mm}
    as Soluble Models }\\
\vspace{5mm}
\vspace{2cm}
{Bruno G. Carneiro da Cunha}\\
{and}\\
{Paulo Teotonio-Sobrinho }\\
\vspace{.50cm}

{\it Universidade de Sao Paulo, Instituto de Fisica-DFMA \\
Caixa Postal 66318, 05315-970, Sao Paulo, SP, Brazil}\\
\vspace{2mm}
\vspace{.2cm}

\end{center}

\begin{abstract}

We study a class of lattice field theories
in two dimensions that includes gauge theories. 
Given a two dimensional orientable 
surface of genus $g$, the 
partition function $Z$ is defined for a triangulation consisting of
$n$ triangles of area $\e$. The reason these models are called 
quasi-topological is that $Z$ depends on  $g$, $n$ and $\e$ but not on
the details of the triangulation. 
They are also soluble in the sense
that the computation of
their partition functions can be  reduced to a
soluble one dimensional problem.  
We show that the continuum limit is well defined if the model
approaches a topological field theory in the zero area limit, i.e., $\e\rt 0$
with finite $n$. We also show that the
universality  classes of such
quasi-topological lattice field theories can be
easily classified. Yang-Mills and generalized Yang-Mills 
theories appear as particular examples of such continuum limits.

\end{abstract}

\newpage

\sxn{Introduction}

Exactly soluble models in statistical mechanics \cite{SOL} and field theory 
are extremely valuable examples
where one hopes to learn about the physics of more realistic
models for which exact calculations are not available. The Ising model,
for instance, has proven to be an incredible source of important 
ideas, such as duality and finite size scaling \cite{BK}, that can be
applied to much more general situations.

The simplest examples of 
soluble models are probably the so called lattice topological field
theories \cite{L,LTFT,LTFT3D,SubDiv}. Let $M$ \input{psbox.tex}
be an oriented 2-dimensional compact
manifold and $T$ a triangulation of $M$.
For instance, the authors of \cite{LTFT}, starting from a quite general 
ansatz, determined what were the conditions on local Boltzmann weights 
such that the partition $Z(T)$ was independent of T. 
In other words, $Z(T)$ was proven to be a topological invariant of $M$.
A large class of models, corresponding to semi-simple
associative algebras, was found. The reason we say that these
lattice theories are soluble is because to compute 
$Z(T)$ for a triangulation $T$
with an arbitrary number $n$ of triangles it is enough to take
another triangulation $T^0$ with the minimal number of triangles
and compute $Z(T^0)$ explicitly. 

Lattice Topological Field Theories (LTFT) are in a sense 
very simple. They are almost
trivial from the point of view of dynamics. Consider for example a
cylinder with 
boundary $S^1\cup S^1$, and the corresponding evolution operator $U$ (or
transfer matrix in the language of statistical systems). It is trivial
to show that for a LTFT the operator $U$  is equal to the identity
when restricted to physical observables
\footnote{However, if
instead of a cylinder one has some other manifold interpolating the
two circles $S^1\cup S^1$, $U$ is no longer the identity.}.
Despite their simplicity, topological models
represent an attractive class of models. They can be generalized 
to higher dimensions and still be exactly soluble. The same type of
models considered in \cite{LTFT} have been carried out in
$3$ dimensions \cite{LTFT3D}. A different approach have
been used by
the authors of \cite{SubDiv} to produce subdivision 
invariant theories in several
dimensions, including four. 

There is a large variety of fully dynamical soluble theories in $d=2$
\cite{SOL}, but in dimensions bigger than 2 this is far from being
true. Unfortunately the general situation is that physical models
in higher dimensions are either soluble but too simple as LTFT's, or
dynamically nontrivial but too hard be be exactly solved, as for example
lattice gauge theories in 3
dimensions. It would be desirable to find a class of models interpolating
these two extreme situations. 
We want to look for models that are a
little more dynamical than LTFT and still can have its partition
function computed. The answer is not known in general, but two
dimensional Yang-Mills theories (YM$_2$) are legitimate examples of
such models.
It is well known
that the partition function of a gauge theory on a 2-manifold $M$ 
is not a topological invariant. Nevertheless its partition function can
be explicitly computed in the continuum and in the lattice
\cite{YM}. It turns
out that the partition function depends not only on the topology of $M$
but also on its area $\a$. Yang-Mills is a deformation of a
topological theory in the sense that it reduces to a topological
theory in the limit $\alpha \rt 0$. This is
an example of what can be called a $2d$ quasi-topological field theory. 
It is well known that $YM_2$ is a particular deformation of a 
topological theory known as $2d$ topological $BF$ theory. There 
are other examples of quasi-topological theories that are deformations
of the same $BF$ theory, where the deformation parameter is again the
area. They correspond to gauge theories known as generalized $YM_2$ \cite{gYM}.

In this paper we shall discuss how to construct quasi-topological
theories on the lattice. They will include gauge theories as a particular
example. Let $M_g$ be an orientable  2d surface  with genus
$g$, and $T(g,n)$ a triangulation of $M_g$ consisting of $n$
triangles. For simplicity, we will
assume that all triangles have the same area $\e$. 
To each link in $T(g,n)$ we associate a dynamical variable
taking values in a discrete (or even continuous) \mbox{set $I$}. Then we follow
\cite{LTFT} and look for models such that the 
partition function $Z(T(g,n),\e)$ depends on the topology through $g$,
on the total number $n$ of triangles, and on $\e$ but not on the
details of the
triangulation $T$. In other words, $Z$ is a function  $Z(g,n,\e)$ of
the topology, the number of triangles and their size. That will
be our definition of a lattice quasi-topological field theory (LQTFT).
As we shall see, 
if the set $I$ of dynamical variables is finite, the partition
function can be exactly computed. In any case, an explicit expression
for $Z(g,n,\e)$ can always be given.
We will
show that the
continuum limit of a LQTFT is well defined whenever the model is
a deformation of a lattice topological theory, i.e, it approaches a
topological field theory in the zero area limit.
The continuum
limit is recovered by taking $n\rt \infty$ and $\e \rt 0$, while
keeping the total area $\a= n \e$ fixed. 

We start by defining a lattice quasi-topological field
theories in Section 2. In Section 3 we compute the partition
function in full generality. The dynamics of LQTFT is discussed in
Section 4. There we compute the evolution operator $U$ for the case of a
cylinder and comment on how to extend the answer to a generic
situation. We also determine what are the physical observables and 
compare with the topological case. In Section 5 we study the
continuum limit. Section 6 is dedicated to a simple example that corresponds to
the generalized $YM_2$. 
Finally on
Section 7, we conclude with some remarks. Some relevant results on  
triangulation of 2d surfaces are given in the Appendix.

\sxn{Quasi-Topological Lattice Theories}\label{S1}

The definition of the model is inspired by  \cite{LTFT}.
Let $T(g,n)$ be a triangulation with $n$ triangles of 
a two dimensional surface
$M_g$ with genus $g$. For simplicity, we will assume that all
triangles have the same area $\e$. A configuration is determined by
assigning to each 
edge of the triangulation a ``color'' $i$ belonging to a index set
$I$. For gauge theories, $I$ is nothing but
the gauge group $G$. To each triangle of area $\e$, with edges
colored by $i,j,k$, we associate a Boltzmann weight $C_{ijk}(\e)$. We assume
that all triangles have the same area $\e$ and that $C_{ijk}(\e)$ is
invariant by cyclic permutation of the color indexes, i.e.,
\be
C_{ijk}(\e)=C_{jki}(\e)=C_{kij}(\e). \label{1.1}
\ee 

An arbitrary triangulation consists of $n$ triangles glued pairwise
along their edges. Therefore it is enough to specify what is the
Boltzmann weight associate with two  joined triangles. Consider the
example in Fig. 1. The corresponding weight is determined by a
gluing operator $g^{kl}=g^{lk}$ and is given by
\be
C_{ija}(\e)~g^{ab}~C_{bkl}(\e),\label{1.2}
\ee
where summations on the repeated indexes $a$ and $b$ are
understood. One may use $g^{ab}$ to lift indexes and write (\ref{1.2})
as ${C_{ij}}^b(\e)C_{bkl}(\e)$ or $C_{ija}(\e){C^a}_{kl}(\e)$. 

\begin{figure}[hbtp]
\begin{center}
\large
\mbox{\psboxto(15cm;0cm){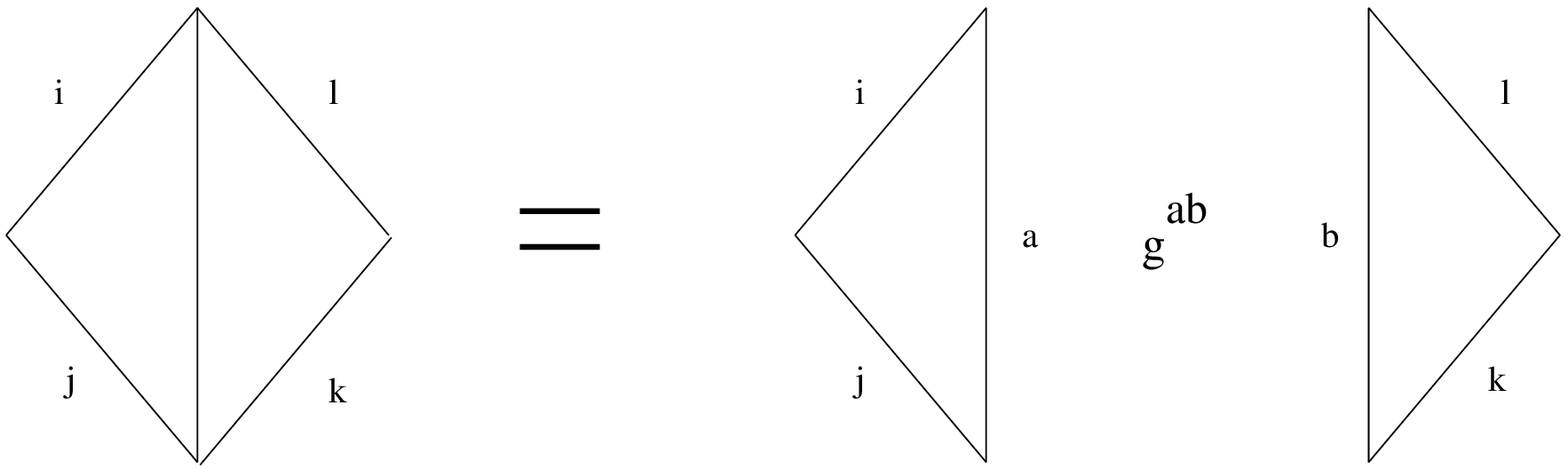}}
\normalsize
\end{center}
{\footnotesize{\bf Fig.1.} The figure shows how the gluing operator
  $g^{ab}$ is used to give the weight corresponding to a pair of glued
  triangles.}
\end{figure}

It will be convenient to restrict the gluing operator $g^{ij}$ in such
way that there exists an inverse $g_{ij}$,
\be
g_{ia}g^{aj}=\d_i^j.
\ee

The partition function for the triangulation $T(g,n)$ is obtained by
performing the gluing operation on all pairs $<ab>$ of edges that
should be identified in order to build the triangulation. 
In other words,
\be
Z(T(g,n),\e)=\prod _{\D\in T}\prod _{<ab>}C_{ijk}(\e)g^{ab}.\label{1.3}
\ee 
If the weights $C_{ijk}(\e)$ are not restricted,
the partition function (\ref{1.3}) depends on the triangulation and 
it can be a complicated and fully dynamical theory. 

It is convenient to represent a given triangulation $T(g,n)$ by its dual
graph $\G(g,n)$. 
Fig. 2 (a) shows the gluing of triangles in terms
of the dual graphs. For an arbitrary graph, such as the one in Fig. 2
(b), one iterates the gluing on its trivalent elementary pieces. The resulting
weight associated with $\G(g,n)$
will be a number that depends on the variables $i_1,i_2,\ldots ,i_6$
attached to the external legs. The graphs must have double
lines in order to encode the same information as the triangulation.
\begin{figure}[hbtp]
\begin{center}
\large
\mbox{\psboxto(15cm;0cm){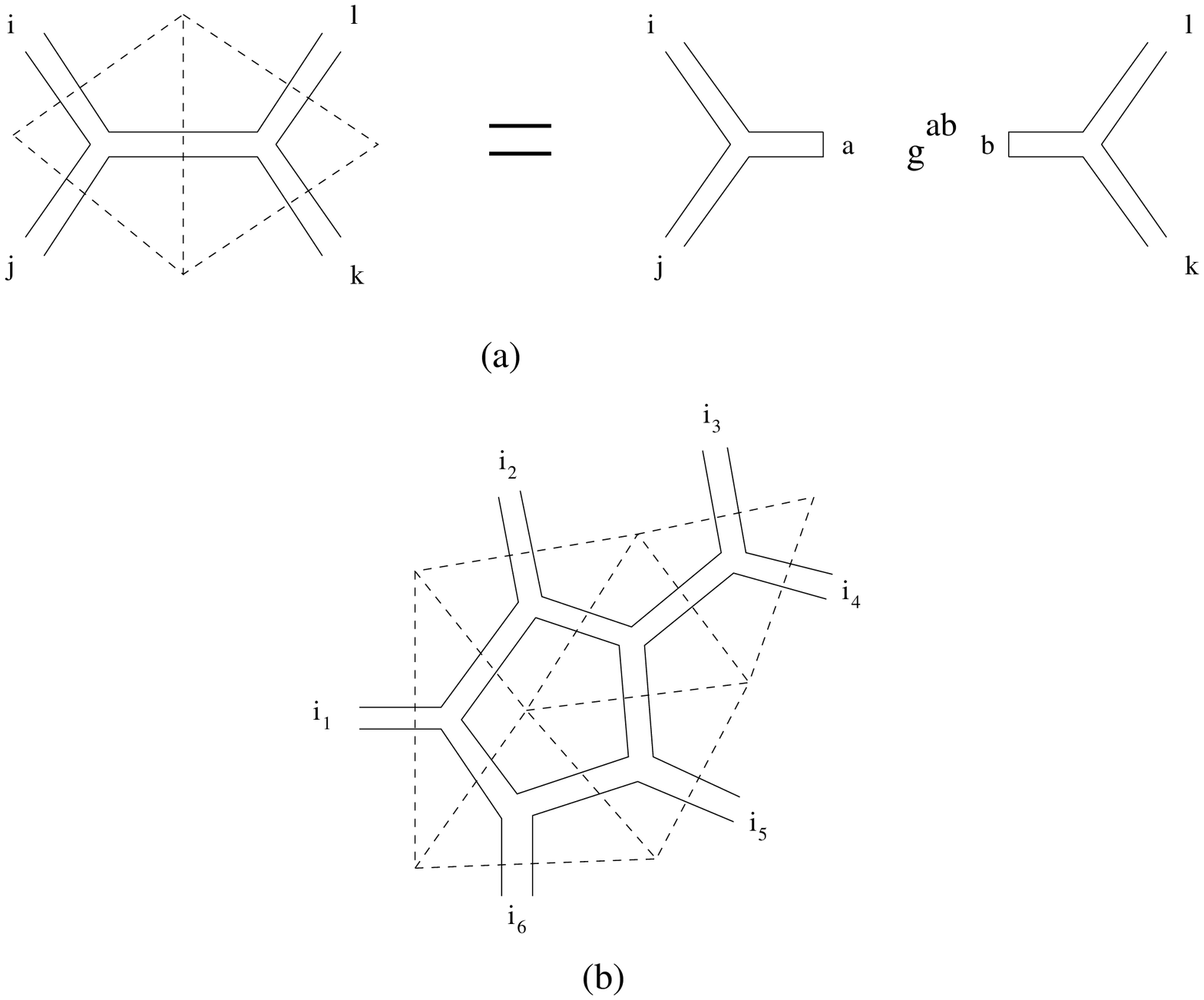}}
\normalsize
\end{center}
{\footnotesize{\bf Fig.2.} Figure (a) shows the gluing of triangles in
  terms of the dual graph. Figure (b) is a simple example of a
  triangulation and its dual graph.}
\end{figure}

Given two triangulations $T(g,n)$ and $T'(g,m)$, or equivalently 
the corresponding 
graphs $\G(g,n)$ and $\G '(g,m)$, of a surface with genus $g$, it is
possible to transform one into another by a
set of local moves that do not change the topology, namely $g$. 
It is well known that two basic moves are needed in order to go from one
triangulation to another. We are going to use the so called flip move
and bubble move. In terms of the dual graphs, these moves are given
in Fig. 3. Note that the flip move preserves the number $n$ of triangles,
whereas the bubble move changes it by two. 
\begin{figure}[hbtp]
\begin{center}
\large
\mbox{\psboxto(15cm;0cm){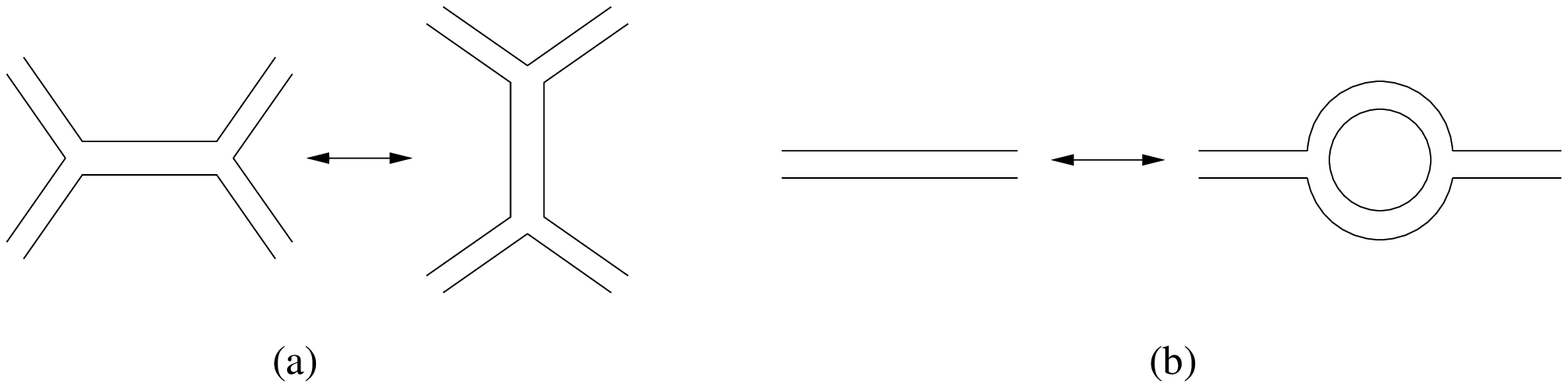}}
\normalsize
\end{center}
{\footnotesize{\bf Fig.3.} The two basic topological moves in terms of
  the dual graphs. Figure (a) is the flip move and figure (b) is the
  bubble move.}
\end{figure}

For a topological theory \cite{LTFT}
$C_{ijk}(\e)$ does not depending on $\e$, i. e.,
\be
\frac{dC_{ijk}(\e)}{d\e}=0
\ee
and it is invariant under both topological moves. 
Invariance under the flip move implies that 
\be
{C_{ij}}^k {C_{kl}}^m = 
         {C_{ik}}^m {C_{jl}}^k,\label{1.4}
\ee
whereas the bubble move is equivalent to 
\be 
C_{iab} {C^{ba}}_j=g _{ij}.\label{1.5}
\ee

A partition function that is invariant under both moves, can not
depend on the triangulation, and therefore it is a topological
invariant of the triangulated surface. 
In other words, $Z$ is a function $Z(g)$ depending only on
the genus $g$ of the surface $M_g$. 

A topological theory defined by $C_{ijk}$ has an
enormous symmetry. Thanks to this fact, the partition function can
be computed. Since $Z$ do not depend on the triangulation, one chooses
the minimal triangulation and writes down $Z(g)$
explicitly. Topological models are very special 
when compared to a generic theory
given by (\ref{1.3}).
In general a model defined by (\ref{1.3}) have  little or no 
symmetry at all. 
What we are going to do is to consider an intermediate situation
where part of the full topological symmetry is not present. That is another
reason for the name lattice quasi-topological field theory (LQTFT). 

The simplest thing to do is to give up the invariance under one of the
two topological
moves described before. It will be interesting to have a
partition function that depends on the size of the lattice, so we choose
to break the invariance under the bubble move and keep the invariance
under the flip move. We also want to 
allow for variation on the size $\e$ of the triangles.
The model is defined by a set of local weights $C_{ijk}(\e)$ invariant
under the flip move, and partition function given by
(\ref{1.3}). In other words, our class of models satisfy the flip move
\be
{C_{ij}}^k(\e) {C_{kl}}^m(\e) = 
         {C_{ik}}^m(\e) {C_{jl}}^k(\e)\label{1.6}
\ee
for any value of the parameter $\e$. 

It may happen that for some critical value $\e=\e_0$, the 
weights  $C_{ijk}(\e)$ also satisfy equation (\ref{1.5}). At this
critical point, the full topological symmetry is restored. As we shall see,
if $\e=0$ is a critical point, the model has a well defined
continuum limit.  

Let us assume for simplicity that the index set $I$ is a finite set
with $r$ elements. 
Consider a vector space $V$ with bases $\{ \f_1,...,\f_r \}$. Then,
for each value of the parameter $\e$ the 
numbers ${C_{ij}}^k(\e)$ define a one parameter family of 
product structures in $V$, namely
\be
\f_i \f_j:={C_{ij}}^k(\e)\f_k.\label{3.4}
\ee
Because of the flip symmetry (\ref{1.6}) the product $\f_i\f_j$
is associative. We may think of ${C_{ij}}^k(\e)$ as given a family
$A_\e$ of algebras on the space of associative algebras defined on $V$.

Since we are  assuming that $g^{ij}$ has an inverse $g_{ij}$
we can define a dual base $\{ \f^i\}$ given by
\be
\f^i=g^{ij}\f_j.
\ee
For the dual basis, the product is
\be
\f^i \f^j:={C^{ij}}_k(\e)\f^k.\label{3.4.1}
\ee

The data we need to define a LQTFT is a pair $(A_\e,g^{ij})$ of
one parameter family of algebras and a bilinear form. There may be some
topological critical point $\e=\e_0$ where equation
$$
C_{iab}(\e_0) {C^{ba}}_j(\e_0)=g _{ij}
$$  
is valid. Such point determines a TFT specified by the data $(A_{\e_0},g^{ij})$.

\sxn{Partition Function}\label{S2}

For a triangulation $T(0,n)$ of the sphere, the corresponding graphs
$\G(0,n)$ are planar. Let $\G(0,n)$ and $\G'(0,n)$ be two planar graphs
representing two different triangulations of $S^2$ but with the same number $n$ of
triangles. It is a well known fact that $\G(0,n)$ and $\G'(0,n)$ can
always be connected via a sequence of flip moves \cite{K}. Therefore if 
$C_{ijk}(\e)$ fulfills equation (\ref{1.6})
the partition function (\ref{1.3}) computed for $\G(0,n)$ and
$\G'(0,n)$ have to be identical.   

Using the same idea of the proof presented in \cite{K} we were able to
show that any pair of dual graphs $\G(g,n)$ and $\G'(g,n)$, for
arbitrary genus $g$, can also be connected by a sequence of flip
moves. For completeness we give a demonstration of this fact on the
Appendix. 
As a result, our partition function (\ref{1.3}) depends only on $g,n$ and
$\e$, provide that (\ref{1.6}) is fulfilled. We will write
$Z=Z(g,n,\e)$ for this matter. The particular graph $\G(g,n)$ used to 
compute $Z$ is immaterial.  

The result of the Appendix shows that any graph $\G(g,n)$ 
can be reduced  via a sequence of flip moves 
to the standard graph $\G^0(g,n)$ given on Fig. 4(a). 
\begin{figure}[hbtp]
\begin{center}
\large
\mbox{\psboxto(15cm;0cm){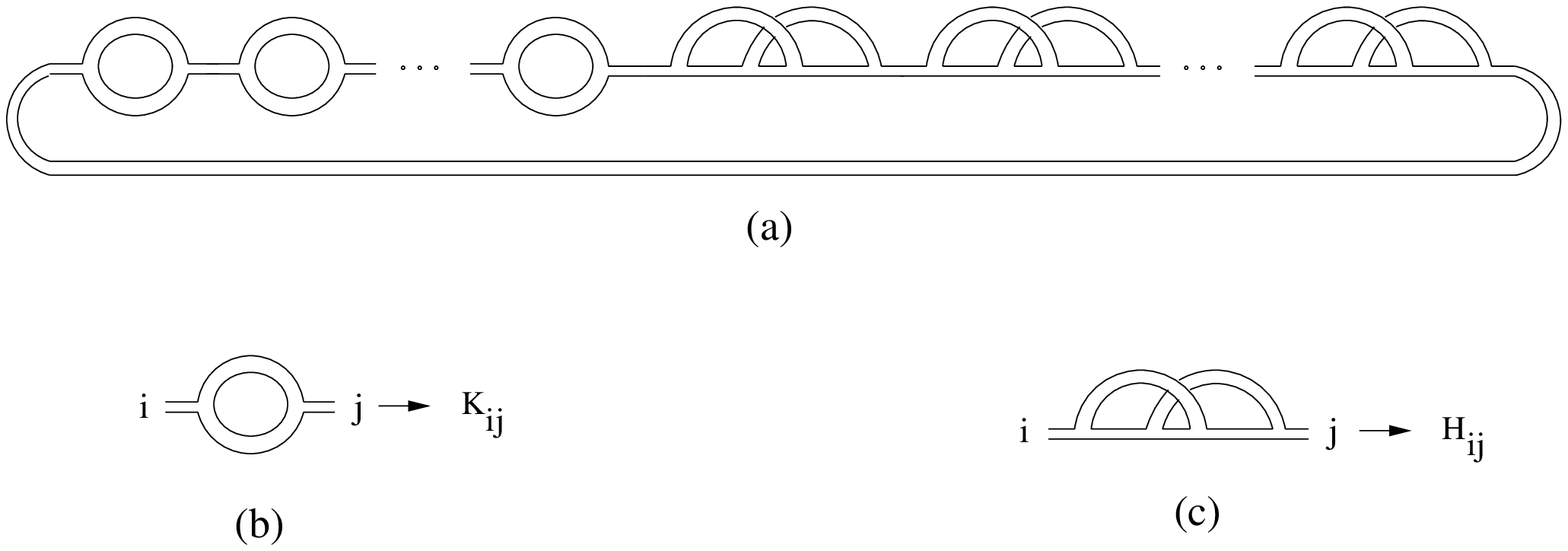}}
\normalsize
\end{center}
{\footnotesize{\bf Fig.4.} The dual graph corresponding to the
  standard triangulation of a surface of
  genus $g$ is given in Figure (a). It can be constructed by repeating 
  the basic blocks shown in Figures (b) and (c) respectively
  $\frac{n-4g}{2}$ and $g$ times.}
\end{figure}

The standard graph is obtained by gluing the elementary blocks
on Fig. 4(b) and \mbox{Fig. 4(c).} From  the gluing rules of last
Section follows that the elementary block of Fig. 4 (b) correspond to
the number $K_{ij}(\e)$ given by
\be
K_{ij}(\e):= C_{iab}(\e)~{C^{ba}}_j(\e)\label{2.1.1}
\ee 
where $ij$ are the variables attached to the external legs.
It is straightforward to verify that
\be
K_{ij}(\e)=K_{ji}(\e).\label{2.1.2}
\ee
Analogously, the elementary block in Fig. 4 (c) gives
\be
H_{ij}(\e):= C_{ikl}(\e)~C^{kmn}(\e)~{C_m}^{pl}(\e)~C_{npj}(\e)
\ee
with 
\be
H_{ij}(\e)=H_{ji}(\e).\label{2.1.3}
\ee

If we define matrices $K(\e)$ and  $H(\e)$ with matrix elements
${[K(\e)]_i}^j=K_{ik}(\e) g^{kj}$ and ${[H(\e)]_i}^j=H_{ik}(\e) g^{kj}$, then the
graph on Fig. 4(a) shows that $Z(g,n,\e)$ can be written as 
\be
Z(g,n,\e)=Tr\left({K(\e)}^{\frac{n-4g}{2}} {H(\e)}^g\right)\label{2.2}
\ee

Equation (\ref{2.2}) and Fig. 4 (a) show that the computation of the
partition function for any 2d lattice surface has been reduced to a one
dimensional problem. If the set $I$ of states is a discrete set with $r$
elements, $K(\e)$ and $H(\e)$ are $r\times r$ matrices. In this case,
(\ref{2.2}) can be calculated for an arbitrary $g$ and $n$.
For this note that the algebra of observables generated by 
$\{ \f_1,...,\f_r \}$
has a natural inner product given by
\be
<\f_i,\f_j>=g_{ij}\label{ip}.
\ee
Using (\ref{2.1.2}) and (\ref{2.1.3}) one can verify that
\begin{eqnarray}
<\f_i,{K_j}^l(\e)\f_l> & = & <{K_i}^l(\e)\f_l,\f_j>;\\
<\f_i,{H_j}^l(\e)\f_l> & = & <{H_i}^l(\e)\f_l,\f_j>.
\end{eqnarray}  
In other words, $K(\e)$ and  $H(\e)$ are self-adjoint with respect to the
inner product (\ref{ip}). Moreover, we will see next that they also commute
\be
K(\e) H(\e) = H(\e) K(\e), \label{2.3}
\ee
therefore they can be simultaneously diagonalized. As the trace is
unchanged by a coordinate transformation, the partition function can
be computed as
\be
Z(g,n,\e)=\sum _{l=1}^{r}{k_l}^{\frac{n-4g}{2}}~{h_l}^g,\label{2.4}
\ee 
where $k_l$ and $h_l$ are the eigenvalues of $K(\e)$ and $H(\e)$. 

We now show that equation (\ref{2.3}) is fulfilled. This is a direct
consequence of the flip symmetry. Consider the graphic representation
of ${K_i}^a(\e)H_{aj}(\e)$ on Fig. 5(a).
\begin{figure}[hbtp]
\begin{center}
\large
\mbox{\psboxto(15cm;0cm){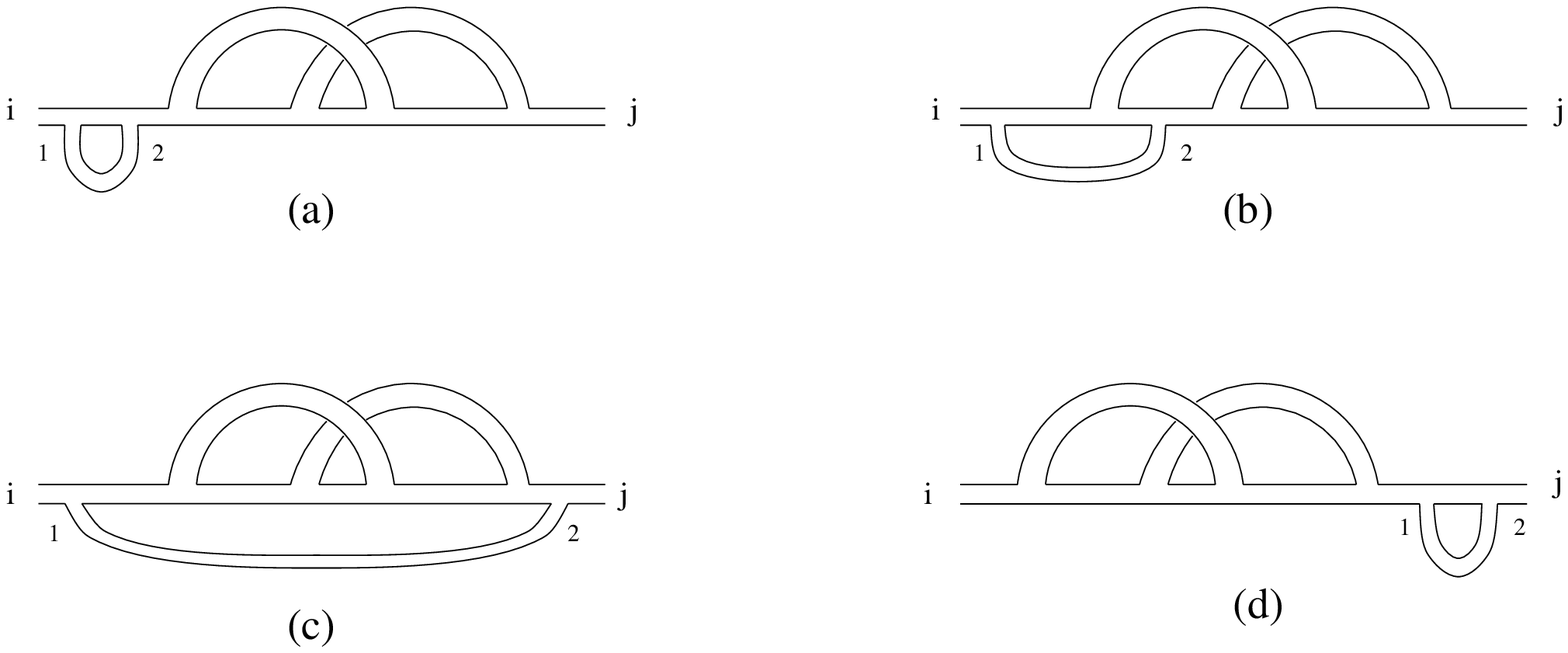}}
\normalsize
\end{center}
\begin{center}
{\footnotesize{\bf Fig.5.} The figure shows equation (\ref{2.5}).  }
\end{center}
\end{figure}
By performing a flip transformation, the leg of
the graph marked with $2$ can be moved to the position presented on
Fig. 5(b). Repeating the same step one can move it further,
arriving at Fig. 5(c). Finally, Fig. 5(d) is obtained by repeating the
process with leg $1$. The interpretation of Fig. 5 (d) in terms of 
${K_i}^j(\e)$ and $H_{ij}(\e)$  reads  
\be
{K_i}^l(\e)H_{lj}(\e)={H_i}^l(\e)K_{lj}(\e).\label{2.5}
\ee  
Therefore $K(\e)$ and $H(\e)$ commute.

\sxn{Dynamics and Observables}

Consider a cylinder where the configuration at the boundary is
fixed. The result of summing over all internal configurations gives us
what we call the evolution operator, or in the language of
statistical mechanics, the iterated transfer matrix. We will denote
the corresponding operator by $U$. 
The dynamical aspects of a model is determined by $U$, since in the continuum
$U$ may be written as the exponential of some Hamiltonian. By its definition, 
a topological theory is such that  $U$ is equal to
the identity when restricted to the physical observables. 
Since our models have less symmetry than a LTFT we
expect that it must have some dynamics.

Let $T(p_1,p_2,n)$ be a triangulation of a cylinder where $n$ is the
number of triangles and  $p_1$, $p_2$ are the number of edges (or
links) on the
boundaries $\s_1$ and $\s_2$. For each boundary circle $\s_1$  and $\s_2$ we
choose a starting link and enumerate the edges in a clockwise
fashion. An example is shown in Fig. 6. 
\begin{figure}[hbtp]
\begin{center}
\large
\mbox{\psboxto(0cm;6cm){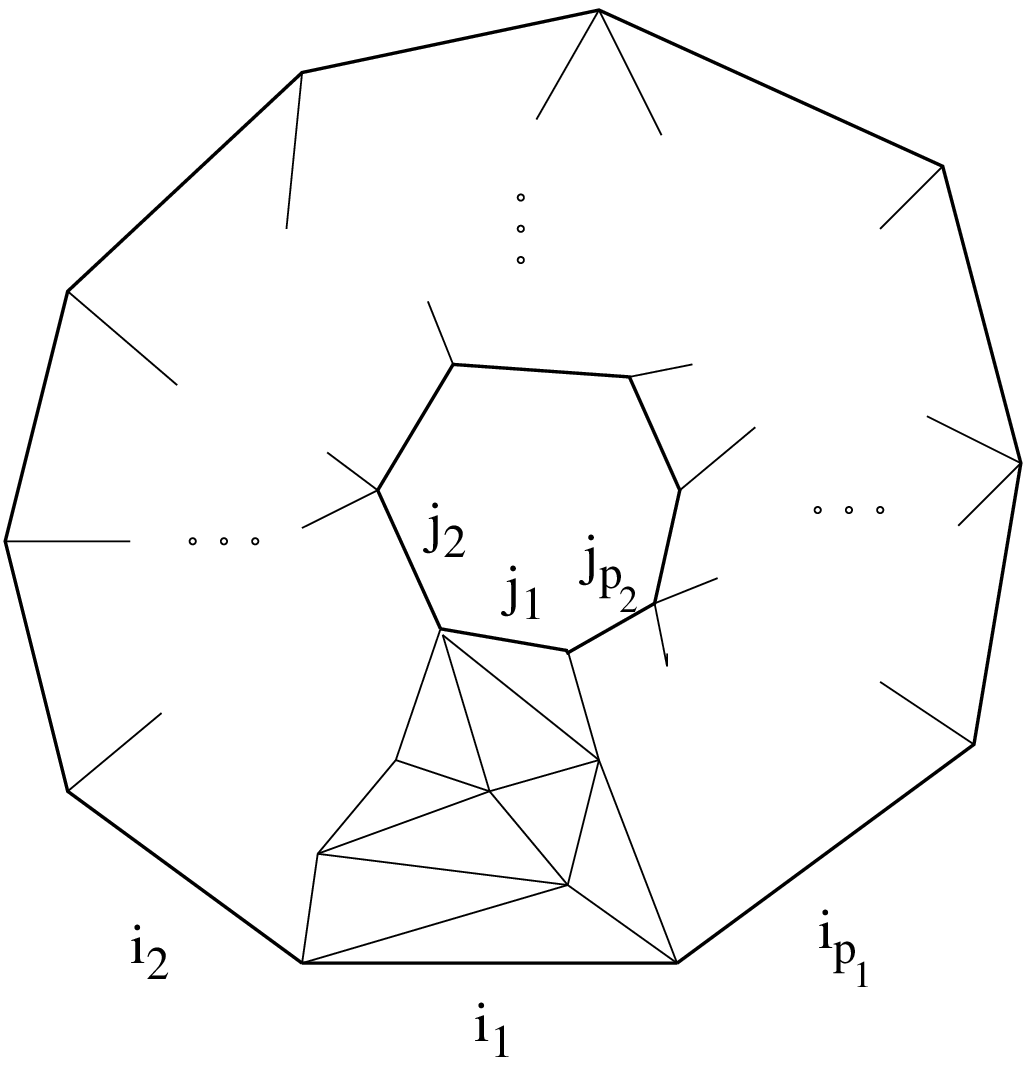}}
\normalsize
\end{center}
{\footnotesize{\bf Fig.6.} A cylinder with boundary given by two
  polygons with $p_1$ and $p_2$ links. The boundary elements are
  enumerated in a clockwise fashion. }
\end{figure}
We define the operator 
$U_{i_1,...,i_{p_1};~j_1, ...,j_{p_2}}$ as the one 
given by the gluing (summing over) of the internal
links according to the rules explained in Section \ref{S1}, while keeping the
boundary configurations on $\s_1$ and $\s_2$  fixed as $(i_1,...,i_{p_1})$ and 
$(j_1, ...,j_{p_2})$ respectively. In other words,
\be
U_{i_1,...,i_{p_1};~j_1, ...,j_{p_2}}=
    \prod _{\D\in T}\prod _{<ab>}C_{ijk}(\e)g^{ab}\label{3.1}
\ee
where $<ab>$ runs over the pairs of glued internal links. What we will call the
evolution operator is the matrix 
\be
U_{i_1,...,i_{p_1}}^{j_1, ...,j_{p_2}}~:=
U_{i_1,...,i_{p_1};~k_1,...,k_{p_2}}
g^{k_1j_1}...~g^{k_{p_2}j_{p_2}}.\label{3.2}
\ee

It is clear from the definition that $U_{i_1,...,i_{p_1}}^{j_1,
  ...,j_{p_2}}$ fulfills the 
factorization properties of an evolution operator. Consider the
splitting of $T(p_1,p_2,n)$ in two cylinders $T_a(p_1,p,n_a)$ and
$T_b(p,p_2,n_b)$, $n_a+n_b=n$, then 
\be
U_{i_1,...,i_{p_1}}^{j_1, ...,j_{p_2}}(T)=
U_{i_1,...,i_{p_1}}^{k_1, ...,k_{p}}(T_a)
              U_{k_1,...,k_{p}}^{j_1, ...,j_{p_2}}(T_b).\label{3.3}
\ee

We are going to assume for the moment that  the set $I$ of variables 
is equal to $\{1,2,...,r\}$. Then, the vector space $V\sim A_\e$ of states 
associated with a single link is generated by a basis $\{ \f_1,\f_2,
...,\f_r\}$. In other words, a generic state $\psi $ is given by 
$\psi = \psi^i\f_i$. 
The space of states $V^{(p_1)}$ corresponding to the boundary
component $\s_1$ with $p_1$
links is just the tensor product 
$V^{(p_1)}= V\otimes V\otimes ... \otimes V$ with $p_1$ factors. At
the boundary $\s_2$ the space of states $V^{(p_2)}$ is defined in the
same way. We recall the
usual interpretation for $U$ as an  
linear operator from $V^{(p_1)}$ to $V^{(p_2)}$ given by
\be
U(\f_{i_1}\otimes ...\otimes \f_{i_{p_1}})=
    U_{i_1,...,i_{p_1}}^{j_1, ...,j_{p_2}}
                   \f_{j_1}\otimes ...\otimes \f_{j_{p_2}}.
\ee

The computation of $U$ follows the same idea as in the calculation of
the partition function in Section \ref{S2}. Given two triangulations 
$T(p_1,p_2,n)$ and $T'(p_1,p_2,n)$ with the same number of triangles,
and the same number of links on the boundary, we were able to show that
they can be connected by a sequence of flip moves. Therefore 
$U$ depends only  on the triangulation through the numbers 
$p_1,p_2$ and $n$. In fact any triangulation $T(p_1,p_2,n)$ can be
brought to  the
standard form given on Fig. 7. The proof is analog to the one
presented on the Appendix section. 
\begin{figure}[hbtp]
\begin{center}
\large
\mbox{\psboxto(15cm;0cm){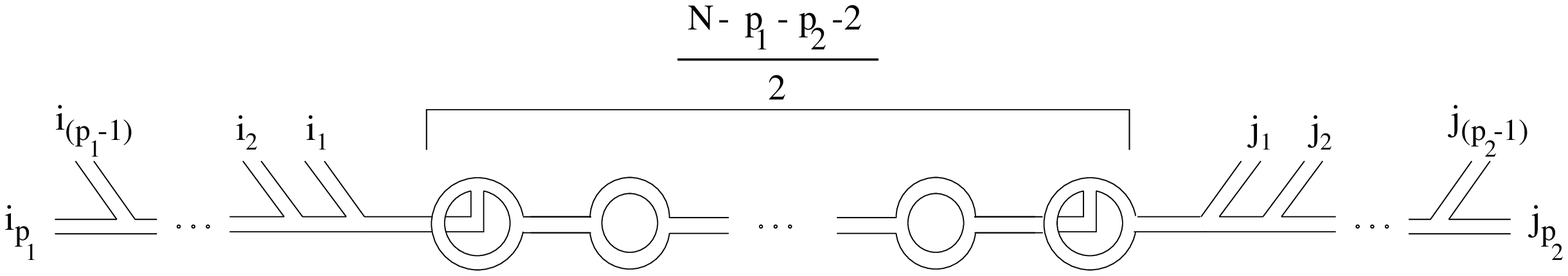}}
\normalsize
\end{center}
{\footnotesize{\bf Fig.7.} The figure shows the standard graph for a
  cylinder. The chain of operators ${K_i}^j(\e)$ starts and ends at a
  new operator ${S_i}^j(\e)$. }
\end{figure}

Note that once more the computation has been reduced
to a one dimensional problem. It involves the product of
the operator ${K_i}^j(\e)$ given in (\ref{2.1.1}), and a new operator 
${S_i}^j(\e)$ defined by
\be
{S_i}^j(\e):=C_{iab}(\e)C^{abj}(\e). \label{3.5}
\ee

We are going to use the following property of ${S_i}^j(\e)$:
\be
{S_i}^m(\e)C_{mjk}(\e)={S_i}^m(\e)C_{mkj}(\e).\label{3.8}
\ee 
A diagrammatic proof of (\ref{3.8}) is given in Fig. 8.
\begin{figure}[hbtp]
\begin{center}
\large
\mbox{\psboxto(10cm;0cm){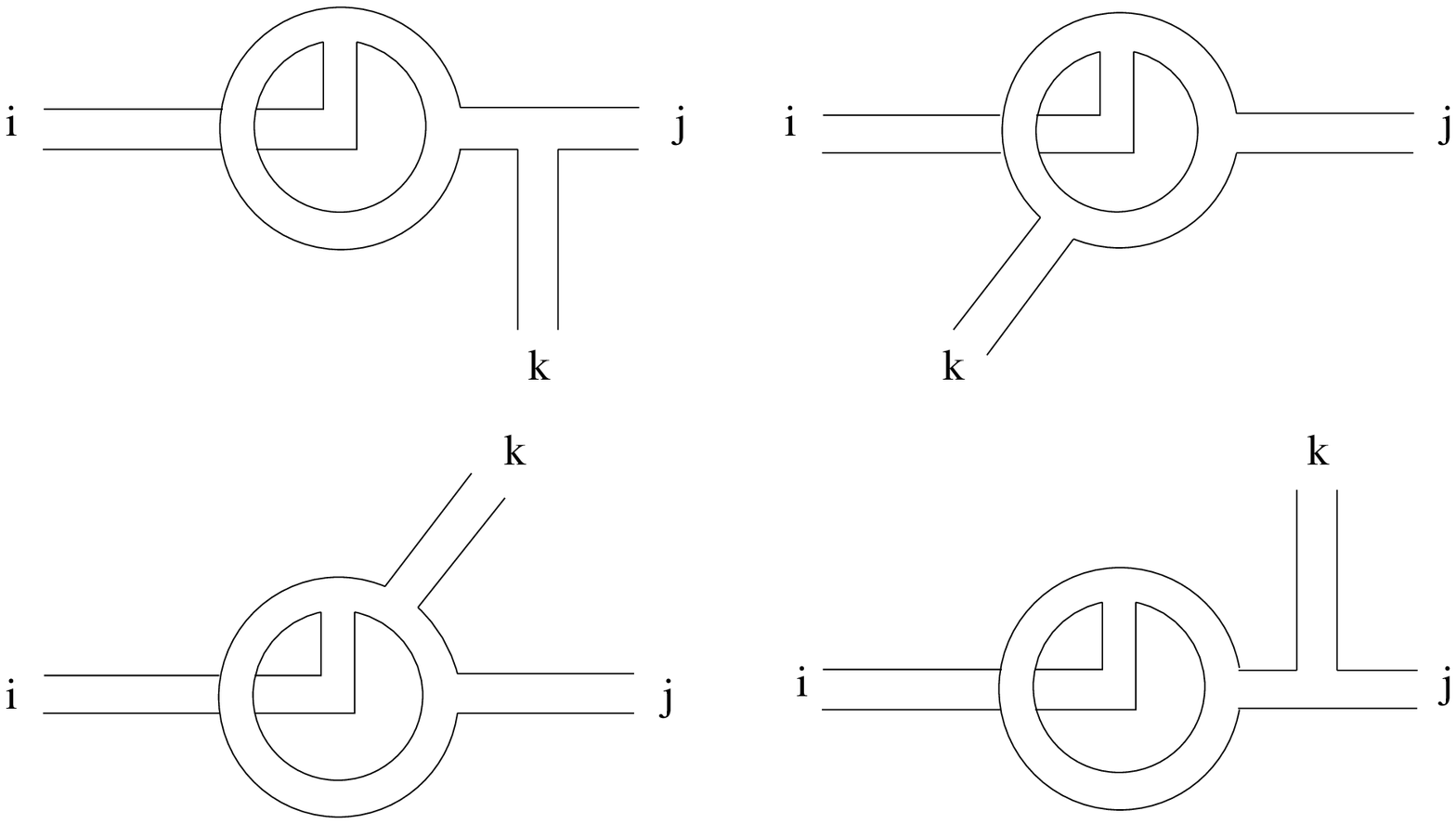}}
\end{center}
\normalsize
\begin{center}
{\footnotesize{\bf Fig.8.} The figure shows a sequence of moves that
  proofs equation (\ref{3.8}). }
\end{center}
\end{figure}

Now let us consider the linear map $S_\e:A_\e \rt A_\e$ given by
$S_i^j(\e)$. On a generic element $a=a^i\f_i\in A_\e$ it act as
\be
S_\e(a)=a^i{S_i}^j(\e)\f_j ~.\label{3.9}
\ee
Using equations (\ref{3.8}) and (\ref{1.1}), it is a simple matter to 
verify that for any $a_1,a_2\in A_\e$ 
\be 
a_1S_\e(a_2)=S_\e(a_2)a_1 \label{3.10}
\ee
and
\be
S_\e(a_1a_2)=S_\e(a_2a_1).\label{3.10.2}
\ee

Equation (\ref{3.10}) shows that  $S_\e(a)$ 
belongs to the center $Z(A_\e)$ of the algebra $A_\e$. In other words,
$S_\e$ maps the algebra $A_\e $ on its center $Z(A_\e)$. The square of
$S_\e$ can also be computed. One can show that
\be
{S_i}^l(\e){S_l}^j(\e)={K_i}^l(\e){S_l}^j(\e)={S_i}^l(\e){K_l}^j(\e).
\ee
Note that in the topological case $K(\e)$ is the identity and $S_\e$ becomes
a projector.

One can see from Fig. 7 that the form of 
$U_{i_1,...,i_{p_1}}^{j_1,  ...,j_{p_2}}(p_1,p_2,n,\e)$  is
\be
U_{i_1,...,i_{p_1}}^{j_1, ...,j_{p_2}}(p_1,p_2,n,\e)=
[S_\e(\f_{k_1}\f_{k_2}...\f_{k_{p}})]^a
\left({K(\e)}^{\frac{n-p_1-p_2-2}{2}}\right)_a^m
[S_\e(\f^{j_{p_2}}\f^{j_{p_2-1}}...\f^{j_1})]_m.
\label{3.5b}
\ee 

The first fact one should notice about (\ref{3.5b}) is that 
$U(p_1,p_2,n,\e)$ is not
an arbitrary operator, but its matrix elements
depend only on $S_\e(\f_{i_1}...\f_{i_{p_1}})$ and 
$S_\e(\f^{j_{p_2}}\f^{j_{p_2-1}}...\f^{j_1})$. Therefore the evolution
given by $U(p_1,p_2,n,\e)$ is actually an evolution for the data
$S_\e(\f_{i_1}...\f_{i_{p_1}})$ belonging to the
center $Z(A_\e)$. Furthermore, equation 
(\ref{3.10.2}) implies that $S_\e$ is invariant under cyclic
permutations
\be
S_\e(\f_{i_1}...\f_{i_{p_1}})=S_\e(\f_{i_{p_1}}\f_{i_1}...\f_{i_{p_1-1}}).
\label{3.5.c}
\ee
For this reason, it becomes useful to introduce loop variables analogue to the
trace of the Wilson loop in gauge theories. Let $\s $ be a loop
on the lattice made of  an oriented sequence $(1,2,...,p_1)$ of
links. Then we define the loop variable $W(\s) \in Z(A_\e)$ as
\be
W(\s)=S_\e(\f_{i_1}...\f_{i_{p_1}}).
\ee
Note that $W(\s)$ depends on the orientation of $\s $, but not on its
starting point.
Analogously, we define its conjugate $\tilde W(\s)=W(-\s)$, where $-\s$
is the same loop with reverse orientation, by
\be
\tilde W(\s)=S_\e(\f^{j_{p_2}}\f^{j_{p_2-1}}...\f^{j_1})\in
            Z(A_\e) \label{3.6}
\ee
The matrix elements of $U$ depend only on $W(\s_1)$ and $\tilde W(\s_2)$, and
therefore in analogy with gauge theories one should regard the loop
variables as the observables of the theory. 

Note that, when restricted to the observables, the evolution $U$ is
given by 
\be
U|_{\rm phy}={K(\e)}^{\frac{n-p_1-p_2-2}{2}}.\label{3.7}
\ee

The observables, or loop variables $W(\s)$, take values in the set
$L(A_\e)$ of elements of the center of the form $S_\e(a)$ 
for some $a\in A_\e$. A natural question to ask is whether $L(A_\e)$ is 
the entire $Z(A_\e)$ or just a subspace. The answer will depend on the
particular set of weights $C_{ijk}(\e)$. However it is possible to
give a sufficient condition such that  $L(A_\e)=Z(A_\e)$. Consider an element
$z=z^i\f_i\in Z(A_\e)$. Using the fact that $z^iC_{ijk}=z^iC_{jik}$
one can show that 
\be
{S_i}^j(\e) z^i={K_i}^j(\e) z^i.
\ee
Therefore if 
$K(\e)$ restricted to the center is invertible then $L(A_\e)$ 
is equal to $Z(A_\e)$. For example this is what happens for the 
topological case when ${K_i}^j={\d_i}^j$.

A Cylinder is topologically equivalent to a sphere with two holes. For
this reason $U$ is also called the two point correlator for genus
zero. To complete our discussion we should consider the corresponding
operator for a surface with $g$ handles and $N$ holes, i.e., the $N$
points correlator for genus $g$. It is a well
known result that it is sufficient to compute the three point
correlator $Y$ for genus zero. Any other correlator can be written in
terms of $Y$ and $U$. Consider a sphere with 3 holes representing a
cobordism from $S^1\times S^1$ to $S^1$. Let 
$T(p_1,p_2,p_3,n)$ be a triangulation with $n$ triangles and
$p_i$ links on the oriented boundary $\s_i$. It is not difficult to show
that  analogously to (\ref{3.5b}) we have
\be
Y_{i_1,...i_{p_1};~j_1,...,j_{p_2}}^{k_1,...,k_{p_3}}(p_1,p_2,p_3,n)=
[W(\s_1)]^a ~ [W(\s_2)]^b ~ {C_{ab}}^l(\e)~ {[K(\e)^\frac{q}{2}]_l}^m 
[\tilde W(\s_3)]_m, \label{3.8a}
\ee 
where $q=n-p_1-p_2-p_3-4$.

\sxn{Continuum Limit}

The continuum limit is obtained by taking  the number $n$ of triangles 
going  to infinity. We will be interested in the scaling situation, when
the area $\e$ of each triangle becomes smaller but the total area $\a$ 
of the surface remains constant. Therefore $\e$ and $n$ are related by
\be
\e=\frac{\a}{n}.\label{4.1}
\ee
At this limit, the partition function  will be a function $Z(g,\a)$ 
of the genus $g$ and the area $\a$.  

Consider the weights associated with the two triangles of
Fig. 9 (a). In the continuum limit $\e$ goes to zero and both
triangles have zero area.
Therefore their weights should be equal. The corresponding diagrams are
shown in Fig. 9 (b). 
It is clear from Fig. 9 (b) that $C_{ijk}(0)$ should satisfy the equation
\be
C_{iab}(0) {C_j}^{ba}(0)=g_{ij}\label{4.2}
\ee
\begin{figure}[hbtp]
\begin{center}
\large
\mbox{\psboxto(12cm;0cm){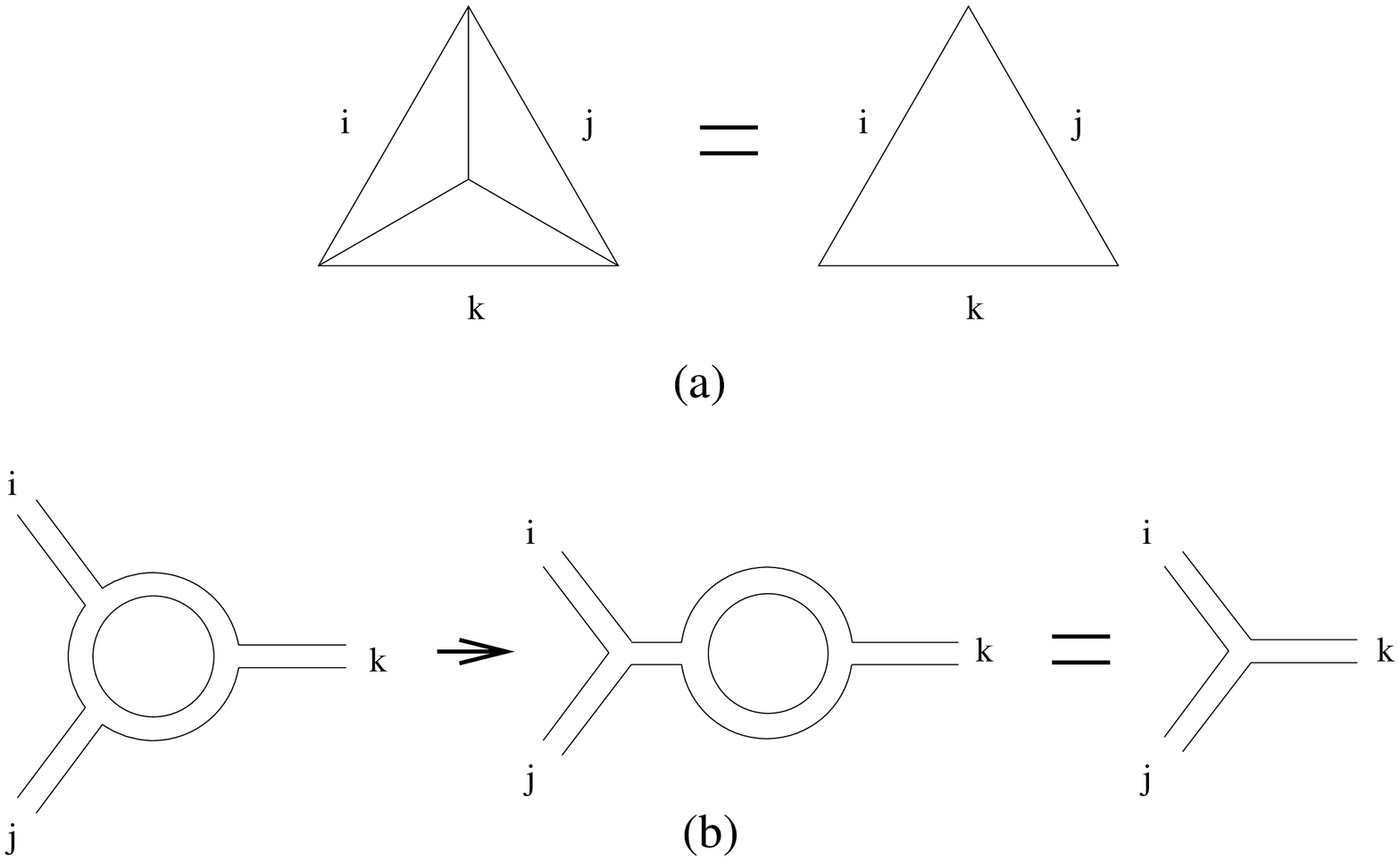}}
\normalsize
\end{center}
{\footnotesize{\bf Fig.9.} In the limit of $\e\rt 0$ both triangles in
  (a) have zero area. The restriction on the weights can be derived from (b). }
\end{figure}
or, in other words ${K_i}^j(0)=\d_i^j$. But (\ref{4.2}) is 
exactly the condition 
(\ref{1.5}) to have a lattice topological field theory. Hence, to have
a well defined continuum limit, the point $\e=0$
has to be a critical point characterized by the algebra $A_0$.
Therefore, to have a well defined continuum limit, the weights
$C_{ijk}(\e)$ have to be a deformation of a LTFT. Therefore
\be
C_{ijk}(\e)=C^{top}_{ijk} +  
  \e \frac{\partial}{\partial \e}C_{ijk}(0) + 
{\cal O}(\e^2).\label{4.3}
\ee
Using (\ref{4.2}) and (\ref{4.3}) in the definition of ${K_i}^j$ we have
\be
{K_i}^j(\e)=\d_i^j+2\e {B_i}^j + {\cal O}(\e^2)\label{4.4}
\ee
where ${B_i}^j$ is defined by
\be
B_i^j:=\left. \frac{1}{2}\frac{\partial }{\partial \e} 
\left( C_{ikl}(\e)C^{lkj}(\e)\right)\right|_{\e=0}.\label{4.5}
\ee
Note that
\be
B_{ij}=B_{ji}\label{5}.
\ee

From (\ref{2.2}) one sees that to compute the partition function  one
has simply to calculate  
${K(\frac{\a}{n})}^{\frac{n-4g}{2}}$ in the
limit ${n\rt\infty}$. Using
(\ref{4.4}) we have
\be
\lim_{n\rt\infty}{K({\frac{\a}{n}})}^{\frac{n-4g}{2}}=
\lim_{n\rt\infty}(I+\frac{2\a}{n}B)^{\frac{n-4g}{2}}=e^{\a B},
\ee
and the partition function is
\be
Z(g,\a)=Tr\left( e^{\a B} H(0)^g \right).\label{4.6}
\ee
Equation (\ref{4.6}) shows that the continuum theory is clearly a
deformation of the TFT characterized by $A_0$. When $\a$ goes to
zero, $Z(g,0)$ becomes topological. 

The operator $U$ also has a well defined continuum limit
when restricted to the physical observable. In the limit $n\rt
\infty$ the algebra $A_\e$ becomes $A_0$.
Let $\s_1$ and $\s_2$ be
the boundary of a cylinder. The observable  are given by two loop variables
$W(\s_1)$ and $W(-\s_2)$ belonging to the center of $Z(A_{0})$. From (\ref{3.7})
we have to compute
\be
U|_{\rm phy}=\lim _{n\rt \infty }{K(\frac{\a}{n}})
      ^{\frac{n-p_1-p_2-2}{2}}.\label{4.7}
\ee
As $p_1$ and $p_2$ are of the order $\sqrt n$, we get  
\be
U|_{\rm phy}=e^{\a B}\label{4.8}
\ee
where $\a$ is the area of the cylinder interpolating between $\s_1$ and
$\s_2$.

It is clear from the above discussion that the continuum
theories are determined by $A_0$ and an operator 
${B_i}^j$.
The algebra $A_0$ defines the topological lattice field
theory that one gets in the zero area limit and and ${B_i}^j$
contributes with a non trivial dynamics. Note that ${B_i}^j$ 
in (\ref{4.5}) is fixed by the derivative of $C_{ijk}(\e)$ at zero. The
global behavior of $C_{ijk}(\e)$ is irrelevant. To classify the
possible continuum theories, or universality classes of a LQTFT,  
one has to determine what are the possible
dynamics ${B_i}^j$ that can come from a generic $C_{ijk}(\e)$ via (\ref{4.5}).
As we shall see, for a given $A_0$, the allowed  
${B_i}^j$ are not arbitrary. 

Let us call $\O(A_0)$ the set of operators ${B_i}^j$ defined by
(\ref{4.5}). Next we show that there is a one to one
correspondence between $\O(A_0)$ and $Z(A_0)$. First we give a
necessary and sufficient condition for a matrix ${B_i}^j$ to be in
$\O(A_0)$ and then we show the correspondence with $Z(A_0)$.

Consider the matrices ${C_m(\e)}$, defined by
\be
{[{C_m(\e)}]_i}^j:={C_{mi}}^j(\e)\label{4.9}
\ee 
One can see from Fig. 10 that ${C_m(\e)}$ fulfills 
\be
{C_m(\e)}K(\e)=K(\e){C_m(\e)}.\label{4.10}
\ee
This equation has to be valid in all orders of  $\e=\frac{\a}{n}$. It is easy
to see that at first order in $\e$, equation  (\ref{4.10}) is equivalent to 
\be
{C_m(0)}B=B{C_m(0)}~~\mbox{ or }~~[B,{C_m(0)}]=0.\label{4.11}
\ee
\begin{figure}[hbtp]
\begin{center}
\large
\mbox{\psboxto(16cm;0cm){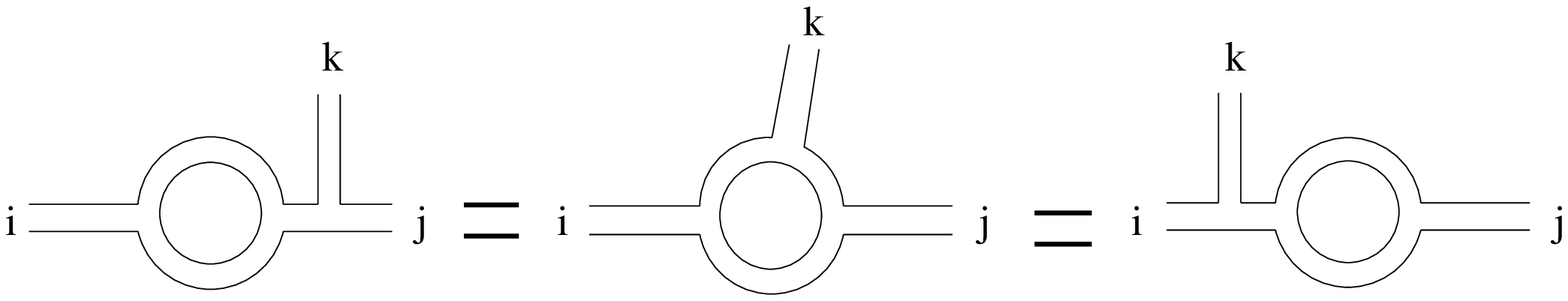}}
\normalsize
\end{center}
\begin{center}
{\footnotesize{\bf Fig.10.} A proof that ${C_m(\e)}K(\e)=K(\e){C_m(\e)}$. }
\end{center}
\end{figure}

Any operator 
${B_i}^j$ coming from (\ref{4.5}) has
to satisfy equations (\ref{5}) and (\ref{4.11}). Actually this is the only
restriction on ${B_i}^j$. Given a topological theory corresponding to
the critical point $A_0$
and an operator ${B_i}^j$ fulfilling (\ref{5}) and (\ref{4.11})
we can always find at least one $C_{ijk}(\e)$ where ${B_i}^j$ comes from. A
simple calculation shows that it is
enough to take
\be
C_{ijk}(\e)=\left[e^{\e B}\right]_i^lC_{ljk}(0).\label{4.12}
\ee  
Therefore, ${B_i}^j \in \O(A_0)$ if and only if it satisfies 
equations (\ref{5}) and (\ref{4.11}). 

Now we can set the correspondence between $\O(A_0)$ and
$Z(A_0)$. Consider the map $\b:Z(A_0)\rt \O(A_0)$ that for any
$z=z^i\f_i\in Z(A_0)$ gives a matrix $\b(z)$ defined by
\be
{{\b(z)}_i}^j=z^m{C_{mi}}^j(0).\label{4.13}
\ee
It is easy to verity that $\b(z)$ fulfills equations (\ref{5}) and
(\ref{4.11}) and therefore it is indeed an element of $\O(A_0)$. The next step
is to find the inverse $\b^{-1}:\O(A_0)\rt Z(A_0)$. Let ${B_i}^j$ be a
matrix in $\O(A_0)$ and define 
\be
\b^{-1}(B)={C_{ia}}^a(0)B^{ij}\phi_j.\label{4.14}
\ee
We have to make sure that $\b^{-1}(B)\in Z(A_0)$. For that consider
the sequence of moves in Fig. 11 (a). Comparing the first and the last
graph we have 
$$
{C_{ij}}^k(0)B^{ib}{C_{ba}}^a(0)={C_{ji}}^k(0)B^{ib}{C_{ba}}^a(0).
$$
Therefore
$\b^{-1}(B)\phi_j=\phi_j\b^{-1}(B)$.

It is very simple to verify that $\b^{-1}\circ \b$ is the identity
map in $Z(A_0)$. Given an element $z^m\f_m\in Z(A_0)$ we will have 
\be
\left[(\b^{-1}\circ \b)(z^m\phi_m)\right]=
   \b^{-1}(z^mC_m)={C_{ia}}^a(0)z^m{C_m}^{il}(0)\phi_l=z^m\phi_m.
\ee
We can also show that $\b\circ \b^{-1}$ is the identity map in
$\O(A_0)$. For that 
consider Fig. 11 (b).
The matrix $B \in \O(A_0)$ is displayed as
a box. It follows from its commutation with $C_m(0) $
that we can attach the box on
any side of the ${C_{ij}}^k$, hence the first step of the figure. Fig.
11 (b) shows that  
${\left[(\b \circ \b^{-1})(B)\right]_i}^j={B_i}^j$, and 
therefore $\b\circ \b^{-1}$ is the identity map.
\begin{figure}[hbtp]
\begin{center}
\large
\mbox{\psboxto(15cm;0cm){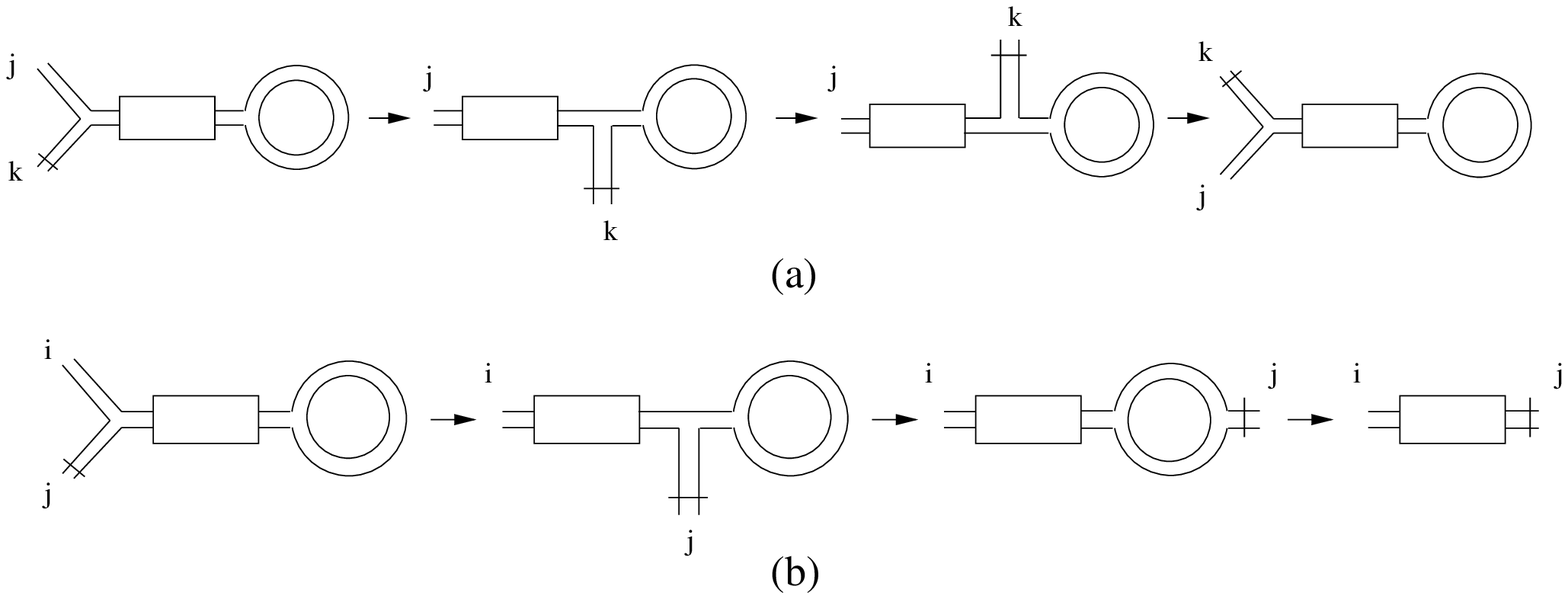}}
\normalsize
\end{center}
{\footnotesize{\bf Fig.11.} Figure (a) shows that the image of $\b$ is
  in the center. Figure (b) shows that $\b\circ \b^{-1} $ is the
  identity map. The crossed strips mean that the corresponding index is
  raised.
}
\end{figure}

In conclusion, the map $\b$ defined in (\ref{4.13}) is a bijection. In
other words all matrices of $\O(A_0)$ are of the form $z^k{C_m(0)}$ for
some $z=z^k\f_k\in Z(A_0)$.

\sxn{Example}

We will now consider an example of quasi-topological theory in the
continuum limit. We
will study the case where the zero area topological theory 
is derived from a group algebra $A_0$.

Given a group $G$, we can construct a group algebra $A_0$ 
over the complex numbers
in the usual way:
$${\bf C}[G]=\bigoplus_{g\in G}{\bf C}\phi_g,$$
\noindent with the algebra product inherited from the group, i.e., 
$\phi_x\phi_y\equiv\phi_{xy}$. 

We can then calculate ${C_{ij}}^k(0)$:
$$
{C_{ij}}^k(0)=\delta (ij,k)~,~~g_{ij}=\delta
(i,j^{-1})~,~~C_{ijk}(0)=\delta (ijk,\I1 ),
$$
where $\I1$ is the identity element in the group.

The next step is to determine $\O(A_0)$, or the set of all
quasi-topological deformation in the continuum. From Section 5, we
know that $\O(A_0)$ is given by all $B_{ij}$ fulfilling (\ref{5}) 
and (\ref{4.11}).
If we denote ${B_i}^j$ by $B(i,j)$, equation (\ref{4.11}) reads
\be
\sum_{l\in G}\delta (ij,l)B(l,k)=\sum_{l\in G}
                       \delta (il,k)B(j,l),\label{6}
\ee
or in other words
$B(ij,k)=B(j,i^{-1}k)$. Therefore 
\be
B(i,j)=B(j^{-1}i,\I1 ).\label{7}
\ee
Using the fact that $B_{ij}=B_{ji}$ together with (\ref{7}) it is easy
to show that
\be
B(i,j)=B(ij^{-1},\I1 ).\label{8}
\ee

Equations (\ref{7}) and (\ref{8}) makes clear that the operator $B(i,j)$ is 
determined by a single function $h:G\rt {\bf C}$ defined as
\be
h(k)=B(k,\I1 ).\label{8.b}
\ee
Furthermore such function satisfies $h(ij)=h(ji)$. Therefore
\be
h(kik^{-1})=h(i)
\ee
and $h(k)$ is a class function, i.e., it depends only on the conjugacy
classes. As a class function, $h(k)$ can be expanded 
on the characters $\chi_R$ of the group. Therefore we can write the operator 
$B(i,j)=h(ij^{-1})$ as
\be
B(i,j)=\sum_RB_Rd_R\chi_R(ij^{-1}),\label{e.1}
\ee
where the sum runs over all irreducible representations of $G$. 
The complex constants
$B_R$ are arbitrary in order to span all possible operators $B(i,j)$.

As we have seen on Section 5 that $\O(A_0)$ is
in one to one correspondence with the center $Z(A_0)$. On the other
hand, equation (\ref{e.1}) shows that $\O(A_0)$ is spanned by the
class functions. Both results are actually equivalent since there is a 
one to one correspondence between the set
of all class functions and the center of the group algebra. 

Each choice of coefficients $B_R$ in (\ref{e.1}) gives us a different
quantum field theory in the continuum. It is interesting to go a
little further and try to identify what could be the Lagrangian
formulation of such field theories. This problem can be solved by
calculating the partition
function for a triangle $\D$ of area $\a$ in the continuum limit.  
For this let us subdivide $\D$ in to 
a triangulation such that the external edges of $\D$ 
are not subdivided. In other words, the corresponding graph has only three
external legs. One can get such
triangulation starting from the one of a cylinder, by closing one of
the boundaries. This correspond to
a particular case of the graph on Fig. 7, where there are only 3
external legs colored by $i,j,k$ on one side an no external legs on the
other side. The resulting graph can be further simplified by applying a
sequence of flip moves and it is equivalent to the one on Fig. 12. 
\begin{figure}[hbtp]
\begin{center}
\mbox{\psannotate{\psboxto(12cm;0cm){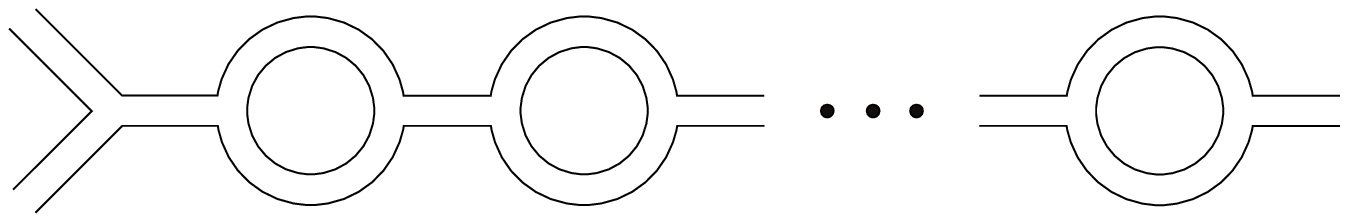}}{ 
\at(0\pscm;0\pscm){$j$}
\at(-0.1\pscm;2\pscm){$i$}
\at(14.3\pscm;1\pscm){$k$} }}
\end{center}
\begin{center}
{\footnotesize{\bf Fig.12.} The standard graph for a triangle with
  finite area.}
\end{center}
\end{figure}
The evaluation of the chain of operators ${K_i}^j(\e)$ gives the
exponential of $B$. The result 
\be
{(e^{\alpha B})_{i}}^{j}=\sum_Re^{\alpha B_R}d_R\chi_R(ij^{-1})
\ee
can be easily computed using using (\ref{e.1}) and the
orthogonality of the characters. From Fig. 12, follows that 
the partition function for the triangle at this limit is:
\be
Z(i,j,k,A)=C^0_{ijl}{(e^{\a B})_{k}}^{l}=\sum_Re^{\a B_R}d_R\chi_R(ijk).
\label{pf}
\ee

Consider a particular case of (\ref{pf}) where $G$ is a Lie group and
$B_R$ is equal to the quadratic Casimir operator $C_2(R)$. The reader 
will recognize the heat kernel action for a triangular plaquette 
that gives Yang-Mills theory in the continuum. This fact allow us to
identify the Lagrangian formulation for this particular quasi-topological
theory as being 2d Yang-Mills. If $B_R$ is not equal to $C_2(R)$
the identification problem becomes less clear.
For such choices of the
coefficients $B_R$ the corresponding theories are not $YM_2$ but
they are still deformations of the same topological theory with the
area being the deformation parameter. 
All one has to do is to find a Lagrangian field theory that gives
(\ref{pf}) as the partition function for a triangle. Such field
theories exist and are called generalized $YM_2$.
We refer to \cite{gYM} for the relevant computations and further details. 

\sxn{Concluding Remarks}

Two dimensional  lattice quasi-topological field theories are ``less
trivial''  than topological models in the sense that they have 
nontrivial dynamics. On the other hand they have less symmetry than the
topological models. In the continuum limit the invariance under the
whole diffeomorphism
group is broken down to the area preserving
diffeomorphisms. Nevertheless the residual symmetry is still enough to
make the model soluble as it allows to reduce a two dimensional
problem to a one dimensional computation. 
If the link variables assume values in a finite
dimensional set, the partition function can be exactly computed. 

The set of  Boltzmann weight ${C_{ij}}^k(\e)$ and the gluing operator
$g^{ij}$ give a one parameter family of associative algebras $A_\e$
together with a bilinear form. We refer to this data as a pair
$(A_\e,g^{ij})$. 
The scaling limit $\e\rt 0 \mbox{, } n=\frac{\a}{\e}$ is
well defined whenever ${C_{ij}}^k(0)$ and $g^{ij}$,  define a lattice
topological field theory.
At $\e=0$ the topological symmetry is
restored and the theory becomes invariant by subdivision. The
continuum theory is not topological and  the partition
function depends also on the total area of the surface. As it is
required by consistence, the theories become topological 
in the zero area limit. This is what is
meant by a quasi-topological field theory, the prototype being 
\mbox{YM$_2$ \cite{YM}.} We have seen that  a single topological theory
given by $A_0$
can be the zero area limit of
more that one continuum quasi-topological theory. The particular
continuum limit will depend on how ${C_{ij}}^k(\e)$
approaches the critical point $\e=0$. 
This is measured by the operator ${B_i}^j$
defined in (\ref{4.5}). We have seen that the set of all
quasi-topological theories associated with $A_0$
is in one to one correspondence to the center $Z(A_0)$ of the
semi-simple algebra $A_0$.

It is not clear which quasi-topological theories in the  continuum  
can be described by means of a Lagrangian. The Lagrangian approach is
certainly possible in the case of YM$_2$ and generalized $YM_2$. 
It would be very interesting to find other examples
of such Lagrangian theories. The simplest solution is to look for 
the analog of a Schwarz type topological field theory, or in other words, 
actions that are invariant under area preserving 
diffeomorphisms. If there are no anomalies, the zero area limit should be a
Schwarz type topological field theory. Volume preserving theories have
been considered in \cite{RB}.
However this may not be 
generic enough and one may need to find quasi-topological theories
that in the zero area limit reduces to Witten's type topological 
field theories . This possibility is presently under investigation and
results will be reported elsewhere.

\appendix

\sxn{Appendix}\label{FM}

We will now present a proof that all triangulations of a given genus
$g>0$ surface  
consisting of $n$ triangles can be 
connected by a sequence of flip moves.
This will be done by an argument somewhat similar to the one
of \cite{K}. The idea is to reduce any triangulation to a special
one which we will refer as standard triangulation. It
consists of several bubble-like
structures  
composed of two triangles, as
well as some  double-handled structures composed of four triangles.
The double handled structures give information about the genus.
The dual graph of a standard triangulation
and the two basic building blocks are shown in Fig. 4. 

It is well known that any surface of genus $g$ can be represented as 
a $4g$-sided polygon with its sides identified suitably \cite{S}. Let
us enumerate  the sides of the polygon as 
$a_1,b_1,a_1^{-1},b_1^{-1},...,a_g,b_g,a_g^{-1},b_g^{-1}$. 
The $g$ genus surface is recovered by gluing
the sides $a_i$ and $b_i$ with $a_i^{-1}$ and $b_i^{-1}$ respectively.

There is a natural distinction among the triangles of the triangulated 
$4g$-sided polygon. 
We will call the triangles external or internal according with whether
they share or not an edge with the boundary of the polygon. Such a distinction 
disappears after the sides of the polygon are identified. 
In this way many different triangulations of the polygon can give the same
triangulation of the surface. The minimum
number of external  triangles is clearly $4g$.

As a first step towards the standard triangulation we will show that no matter
how complicated is the triangulation of the surface, we can always
reduce the number of external triangles per side of the polygon by one and
therefore reduce it to the minimal number. In other words, 
it is enough to consider triangulated polygons with only $4g$ external
triangles. Suppose that there is a side of the  polygon consisting of $l$
links. Consider any two consecutive links, say $\bar{AB}$ and
$\bar{BC}$, and their respective triangles. Observe that
one can always perform a sequence of flip moves in order to make these
two triangles share an edge. Now we follow the steps shown in Fig. 13.  
We can flip the common edge and as a
result the two  
consecutive links now belong to the same triangle $ABC$. 
To proceed, remember that this
polygon is in fact a genus $g$ surface, and therefore the side
containing $\bar{ABC}$ is identified with another side. Clearly
the notion of  
which sequence of links makes the identified sides is somewhat
arbitrary. The links  
defined as the side of the polygon could be in fact any edge nearby. In
particular, we can replace $\bar{ABC}$ by $\bar{AC}$.
In this way we arrive at the last picture of Fig. 13. The
number of links in a given side of the polygon has been reduced by
one. We get a triangulated polygon with $4g$ external triangles by
iterating the process. 
\begin{figure}[hbtp]
\begin{center}
\large
\mbox{\psboxto(15cm;0cm){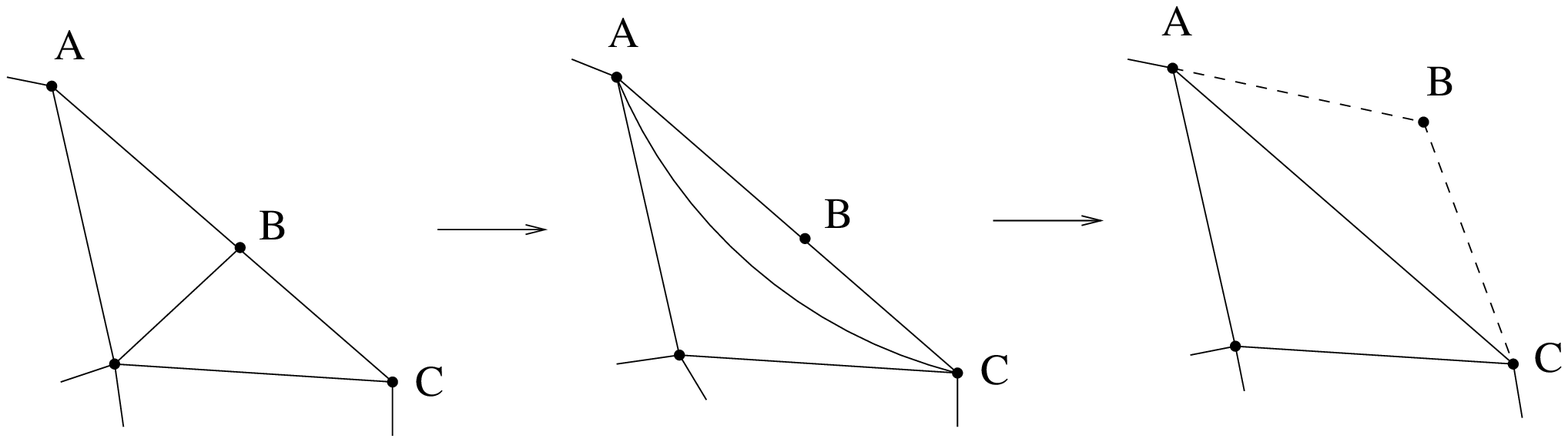}}
\normalsize
\end{center}
{\footnotesize{\bf Fig.13.} How we can decrease the number of
  external triangles. From left to right: in the first figure we
  have two triangles, and the external edges are $\bar{ABC}$, in the
  second we made a flip move, in the third we redefined our external 
  edge as $\bar{AC}$. The dashed line means that $B$ has been sent
  ``to the other side''. }
\end{figure}

The dual graphs of triangulated polygons are planar. Therefore we
can simplify the  
pictures by using single lines instead of double lines when drawing
them. The  
$4g$ external triangles will be represented by $4g$ external legs in
the dual graph numbered 
by $a_1,b_1,a_1^{-1},b_1^{-1},...,a_g,b_g,a_g^{-1},b_g^{-1}$.
It is not difficult to see what is the general structure of the dual
graphs. After considering some 
examples, like the one in Fig. 2, one realizes that the graphs
consist of a big external with the $4g$ external legs attached to it
plus several internal loops all interconnected. We now follow
\cite{K} to arrive at the standard triangulation for the $4g$-sided
polygon.

One should remember that 
the action of the flip move is simple the sliding of one edge over
another as shown on Fig. 14.
\begin{figure}[hbtp]
\begin{center}
\large
\mbox{\psboxto(15cm;0cm){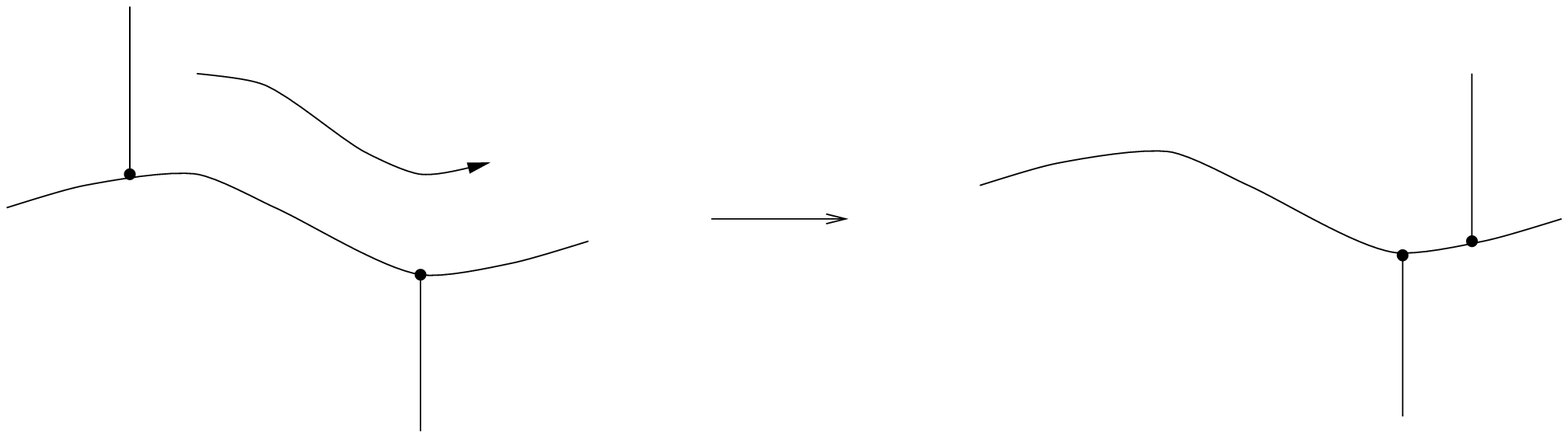}}
\normalsize
\end{center}
{\footnotesize{\bf Fig.14.} The action of the flip move on dual
  graphs. As the graph is planar, we can represent it by single lines.
  The figure shows the flip move as the sliding of lines one over another.}
\end{figure}

Consider now the internal loops. By sliding
one line over another, we can arrange all lines connecting the several
loops in such way that a given loop  is linked to at most two different loops.
Fig. 15 shows how to
disentangle compounds of this type, forming ``pins'' in the process.   
\begin{figure}[hbtp]
\begin{center}
\large
\mbox{\psboxto(15cm;0cm){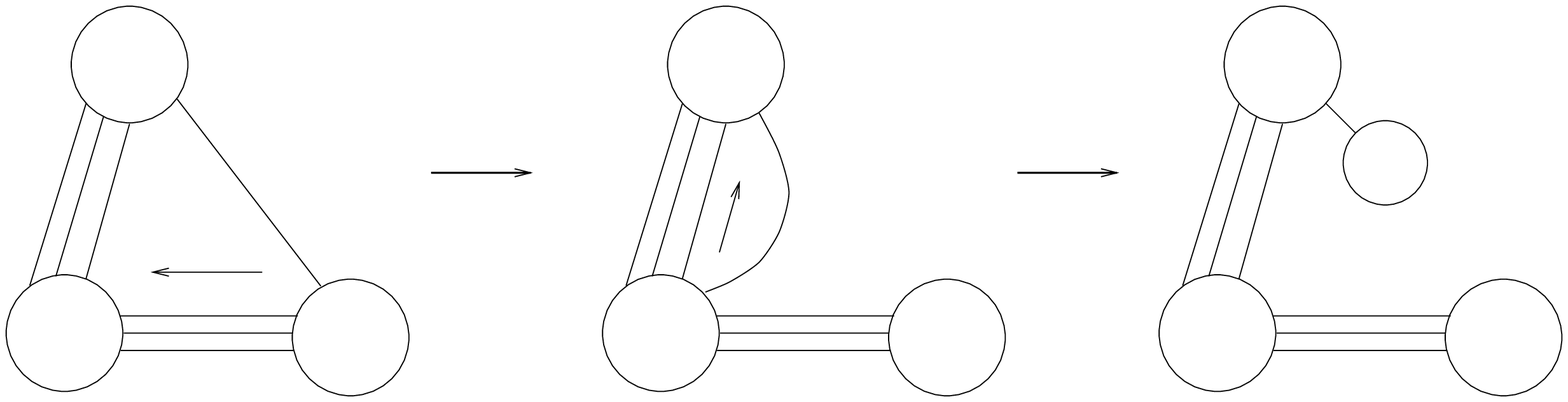}}
\normalsize
\end{center}
{\footnotesize{\bf Fig.15.} How to disentangle interconnected loops. 
This results in a number of   ``pins'' attached to any of them.}
\end{figure}
We are left then with chain-like structures of loops linked by some
various numbers of lines. \mbox{Fig. 16} shows how to reduce the number of
lines to just one line, again forming pins.
\begin{figure}[hbtp]
\begin{center}
\large
\mbox{\psboxto(15cm;0cm){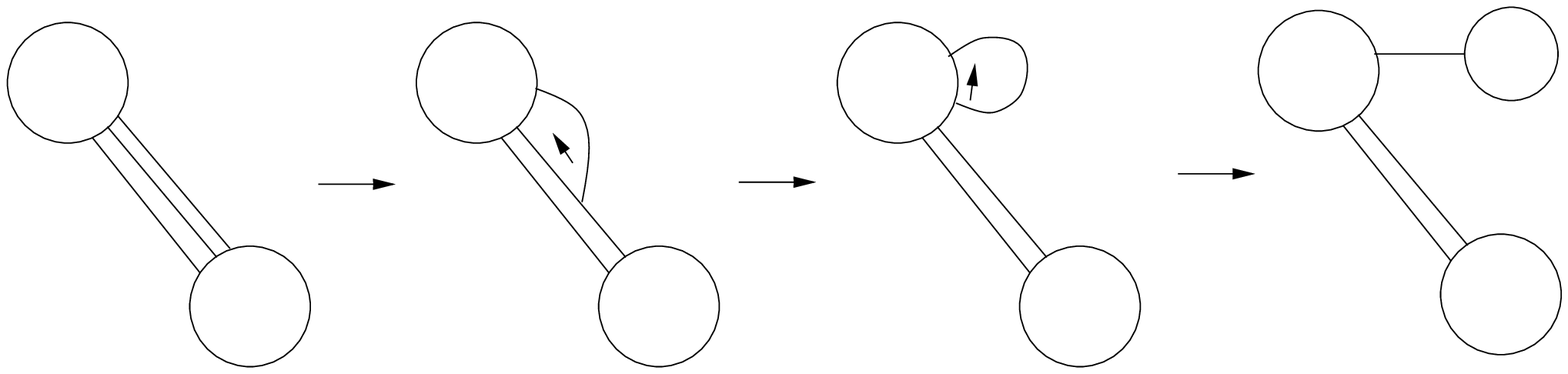}}
\normalsize
\end{center}
{\footnotesize{\bf Fig.16.} How to reduce the number of lines
  connecting two loops creating ``pins''.}
\end{figure}
We have by now some internal loops, with some pins attached,
connected by just one line to at most two different loops, and this compound
connected by just one line to a greater external loop. 
The pins can be carried one by one to the last
internal loop -- the only loop that is connected to just one different
loop -- and thus becoming the last one. Repeating this process, we will
eventually reach a triangulation with a chain of loops linked by only one
line to each other. One end of the chain will be linked from inside
with the large  external loop. After gluing the external legs labeled 
$a_i$ and $b_i$ with $a_i^{-1}$ and $b_i^{-1}$ one gets graph
equivalent to the one on Fig. 4 (a). 

One may notice that, although all we did was for a surface with genus
greater than zero, we could also extend the argument for genus
zero. However, this is exactly what is done in \cite{K} and we shall
not repeat the argument here.

\newpage

\noindent
{\large \bf Acknowledgments}\\
This work was supported by CNPq.

\end{document}